# Multiple String LED Driver with Flexible and High Performance PWM Dimming Control

M. Tahan, Student *Member, IEEE*, T. Hu, *Senior, IEEE*

*Abstract*— The main objectives in driving multiple LED strings include achieving uniform current control and high performance PWM dimming for all strings. This work proposes a new multiple string LED driver to achieve not only current balance, but also flexible and wide range PWM dimming ratio for each string. A compact single-inductor multiple-output topology is adopted in the driver, accompanied by synchronous integrators and variable dimming frequency, to achieve both high efficiency and high performance dimming. By using the proposed variable dimming frequency scheme, high dimming frequency is applied to a string with high dimming ratio, which helps to maintain the deviation of LED string current in an acceptable range, while low dimming frequency is applied to a string with low dimming ratio, which helps to achieve rectangular LED current waveform. Meanwhile, the new time multiplexing control scheme automatically optimizes the LED strings' bus voltages, thus minimizes each string's power loss. A three string LED driver prototype is constructed to validate the effectiveness of the proposed control scheme, where the three strings can have different dimming ratios between 4% and 100%.

## I. INTRODUCTION

LIGHTING Emitting Diodes (LEDs) are receiving more and more attention because of their benefits such as longevity, chromatic variety, and fast response. One of the main objectives in LED driver design is to minimize color spectrum shift and to control luminance intensity via accurate current regulation. The luminosity of the LEDs is directly related to the forward current, while their I-V characteristics are temperature dependent and may vary with aging [1, 2]. Due to the exponential relationship between the forward current and the forward voltage of LEDs, a small voltage variation may cause dramatic increase of the forward current. Since the forward voltage varies in a wide range even for LEDs from the same manufacturer [3, 4], most LED drivers have been developed to control the illumination by regulating the forward current to a desired value. In a configuration with parallel strings, current balance mechanism is usually utilized to achieve uniform brightness.

In view of these concerns, advanced control strategies have been developed, along with complex driver configurations. In conventional balancing method a resistor is placed in series with each string to minimize the current difference [5, 6]. However, the series current sense resistor may cause significant power loss and poor efficiency. To reduce such a power loss, reactive balancing method is proposed in [7-9], and [10]. Resonant capacitors in series with the common-mode choke and a two output rectifier structure is proposed in [9], where the currents of the two outputs can be automatically balanced due to the charge balancing of the resonant capacitors. In order to extend this configuration to *2n* outputs, *n* separated transformers with all primary side windings connected in series are needed [11]. In [10], a lossless reactive component drives the LEDs by an AC source coupled with a diode rectifier. The impedance of the capacitors is high enough and the tolerance is in a small range to determine the current and restrict the current difference among LED strings. However, the utilization rate of LEDs is relatively poor, and the pulsating current in LEDs may decrease the luminous efficacy [12]. Dedicated dc-dc converters are used in [13, 14] to reduce the power dissipation, where each converter is designed and controlled optimally to achieve desired forward current. A drawback with this method is that the configuration is complex and needs many bulky magnetic components, which considerably increases the size and the construction cost. In [15], a digital PI controller combined with a linear current regulator is used to regulate the boost converter voltage and achieve desired current for each LED string. To equalize the current in each LED string, [16] and [17] utilize linear current mirror (CM) regulator. In CM method, LED strings' currents are balanced based on one controlled current source as a fixed current reference, which requires a separate power supply. The voltage difference between forced operating voltage by transistor and supply voltage, results in a high headroom voltage and serious power dissipation. A time multiplexing control scheme along with high-frequency-time-sharing (HFTS) PWM dimming technique is presented in [4], where most of the effort is devoted to minimizing power loss and the volume of the driver. By using the control scheme in[4], the average current in each LED is a function of the number of LED strings (N),

$$I_{avg} = V_{ref}/(N \times R_{led}) \qquad (1)$$

Manuscript received Sept. 12, 2016; revised Dec. 05, 2016; accepted Jan. 06, 2017. Date of current version Jan. 18, 2017. This work was supported by NSF under Grant ECCS-1200152.

Mohammad Tahan is a PhD student in the Department of Electrical and Computer Engineering, University of Massachusetts Lowell, Lowell, MA 01854, USA (e-mail: Mohammad_tahan@student.uml.edu).

Tingshu Hu is with the Department of Electrical and Computer Engineering, University of Massachusetts Lowell, Lowell, MA 01854, USA (e-mail: Tingshu_hu@uml.edu).

Which results in a decrease of the utilization factor. In other words, the number of LED strings is limited by $N_{max} = I_{PK}/I_{led}$, where $I_{PK}$ is the maximum peak current of an LED device. To overcome the limitations of existing SIMO LED drivers, a coordinated low-frequency and time-sharing (CLFTS) technique has been proposed in [18] and [19]. By using the CLFTS technique, the low-frequency dimming switches lead to a non-pulsating output current profile during the dimming-ON period, while the high-frequency time-sharing operation of the power channel switches enables a full range dimming ratio for every string. With this new technique, LED devices of low current ratings can be utilized. It should be noted that the LED driver developed in [18-20] is based on a SIMO buck converter which requires the dc power supply voltage to be higher than the LED strings' channel voltages.

The objective of this paper is to develop a SIMO LED driver based on a boost converter, to achieve flexible and wide range PWM dimming for every string. The main difference from the LED driver in [18] and [19] is that the LED driver in this paper allows the power supply voltage to be in a wide range below the LED strings' channel voltages, making it more applicable to portable devices powered by batteries. Another difference is that a low frequency time-sharing among the power channel switches is implemented in this paper, which allows full range duty ratio for the boost converter's main switch when each channel is charged. This in turn gives full control to the boost converter, which is essential to a wide range of power supply voltage, robust stability and good transient performances. These features of the proposed LED driver is enabled by a unique combination of synchronous integrators, variable dimming frequency and a new time multiplexing scheme.

This paper is organized as follows. In Section II, the operating principle is illustrated, along with a timing diagram for the switches. For simplicity and clarity, a driver for one LED string is utilized to explain the proposed control strategy and then extended to a driver with multiple LED strings. Selection of key components for boost converter and design considerations are discussed in Section III based on transient analysis results. Variable dimming frequency is developed in Section IV to attain high performance. Experimental results are presented for a three LED string driver in Section V, which validate minimal current balance error and high efficiency. Finally, Section VI summarizes the main results and the contributions of the paper.

## II. Operation Principle Of Proposed LED Driver

To achieve optimal LED performance and lifetime expectancy, it is desirable to have a driver design that yields accurate LED currents and minimal power loss. One key challenge in the traditional LED drivers with multiple LED strings is system volume. In this paper, a compact single-inductor multiple-output (SIMO) LED driver based on variable dimming frequency and synchronous integral control scheme will be proposed. For simplicity and clarity, the system architecture of the proposed driver with a single LED string is presented first and described in details. This architecture is then extended to a driver for multiple LED strings, with detailed explanation of the multiplexing timing diagram and variable dimming ratio.

### A. Single LED String Driver

Ideally, maintaining the bus voltage ($V_O$) at a constant desired value leads to a corresponding desired LED current. However, the LED's $I$-$V$ characteristics depends on the temperature and may vary with aging. Hence, it is preferred to control the LED current directly so that it follows a desired reference value. For this purpose, the LED current is sensed by $R_S$ and feedback control is designed to achieve reference tracking. It was shown in [21-23] that a simple integral control can achieve practical global stability for general dc-dc converters. Thus in the proposed control scheme, the integration of the current error is used for feedback, as shown in Fig. 1. In the control loop, the first op-amp picks up the LED current and amplifies the signal in order to raise the signal to noise ratio. The second op-amp generates the tracking error of the current whose output is connected to MOSFET $Q_3$. The source of $Q_3$ is connected to the 3rd op-amp, which is an integrator. The 4th op-amp scales the integrator output so that its lower limit (a negative value) corresponds to the upper limit of the duty cycle $D_C$ for the boost converter. The input of the integrator is turned on and off by $Q_3$. When it is turned off, the voltage across $C_f$ does not change. The two MOSFETs $Q_2$ and $Q_3$ are controlled by the same output of PWM2, so that the LED string and the integrator are turned on and off at the same time. As $Q_2$ is turned on and off by a PWM dimming signal, the LED current that flows through the string is theoretically an ideal rectangular waveform. For driving $Q_1$, the logic AND ensures that the boost converter is turned off while the LEDs are turned off. The LED current depends linearly on $V_O$ around a small region. Although this relationship may vary with time, the change is too slow as compared with the dimming frequency and the transience of the boost converter. In the implementation of a single LED string driver, PWM1 controls the boost converter at a higher frequency about $400kHz$ and PWM2 implements dimming across the LED string at a lower frequency, e.g., $300Hz$. Therefore, by controlling PWM1 via feedback loop, the bus

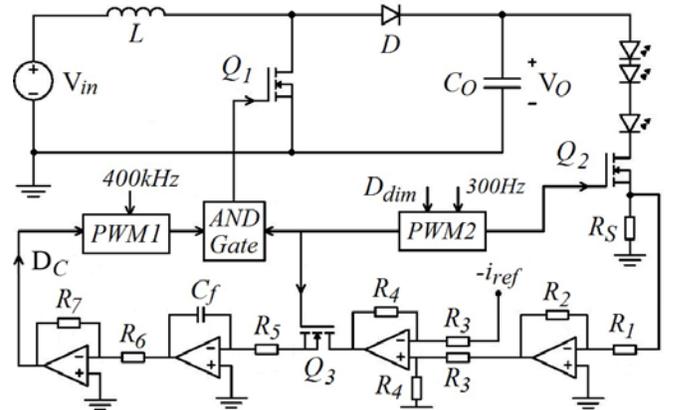

Fig. 1. Single string LED driver with synchronous integrator

voltage ($V_O$) and the LED current are automatically regulated to the desired values.

### B. Multiple LED String Driver

In conventional methods which use linear current regulators to control LED strings' currents, all local bus voltages of LED strings are adjusted to the same value. This may cause excessive power loss in the current regulators, when the difference between bus voltage and LED string's voltage varies from low to high. In this work, the LED current of each string is controlled based on its own *I-V* characteristic, thus independent bus voltage regulation is achieved to prevent excessive power loss. The proposed multiple string LED driver circuit is shown in Fig.2. As compared to the single string driver, an additional switch $Q_f$ is used in parallel with the inductor and one more switch is used for each string ($Q_2$, $Q_5$ and $Q_8$ for the three strings respectively). A dc–dc converter normally works in discontinuous conduction mode (DCM) at light load for high efficiency, while it operates in continuous conduction mode (CCM) at heavy load to supply more power. Although smaller inductor needs to be used for heavy load to operate at DCM to attain higher efficiency, it causes higher peak inductor current, larger current ripple, and consequently larger output voltage ripple. For a SIMO converter, CCM leads to cross regulation among channels and DCM imposes large current stress on the components.

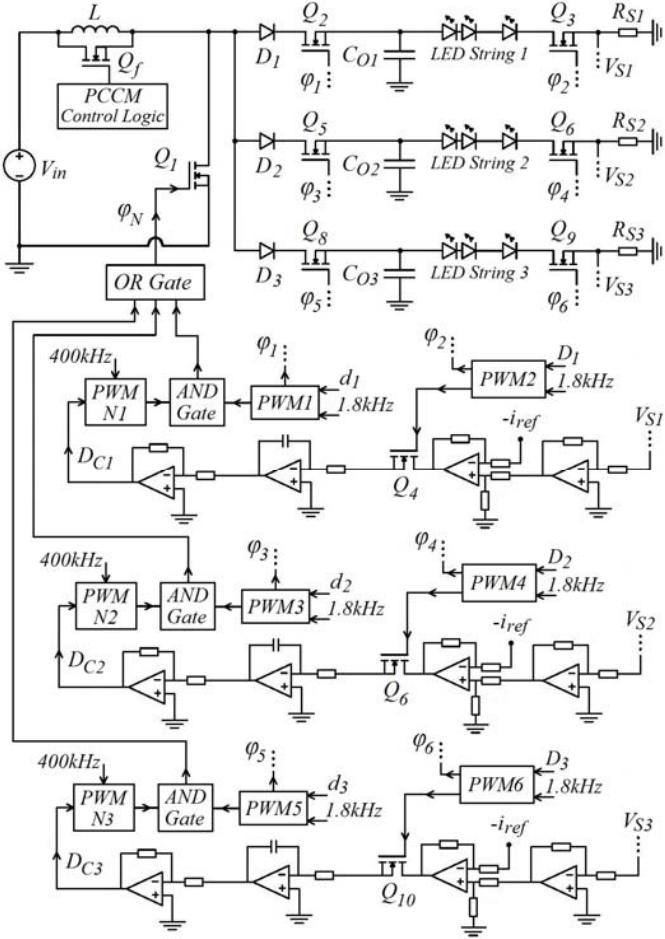

Fig. 2. Circuit block diagram for proposed LED driver

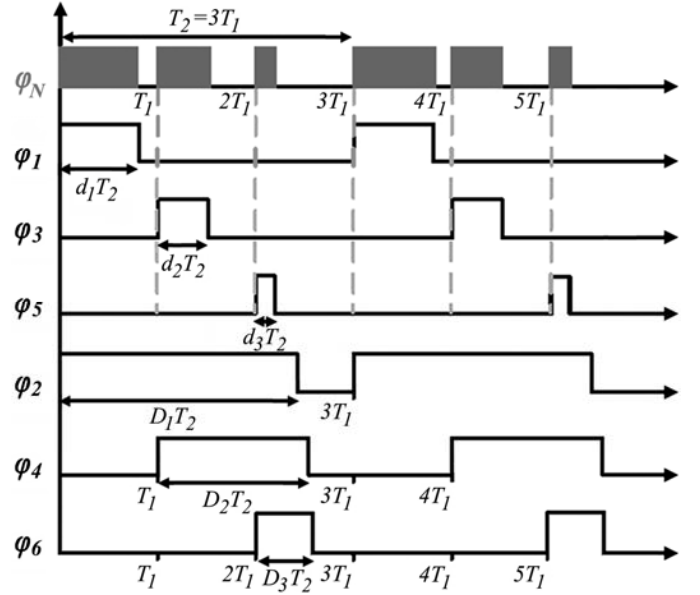

Fig. 3. Timing diagram

The pseudo-continuous conduction mode (PCCM) is proposed for SIMO converter in [24] for cross regulation suppression, as well as for handling large current stress at heavy loads. To implement PCCM, a freewheel switch $Q_f$ is applied in parallel with the inductor, which shorts out the inductor in the DCM interval when the current reaches zero. In this working mode, the floor of the inductor current is raised by a dc level of $I_{dc}$. This eliminates the power constraints in the DCM case, while retaining relatively small current ripple and voltage ripple. With this method, a larger bandwidth is achieved and the transient response is improved [24]. In this work, PCCM is employed for improving transient response and mitigating the voltage stress.

In what follows, the proposed timing scheme for the switches will be described in detail for the case of three parallel LED strings, as illustrated in Fig. 3. At first, assume that a uniform dimming frequency is adopted for all strings and the corresponding dimming period is $T_2$. The dimming ratios for the three strings are denoted as $D_1$, $D_2$ and $D_3$, and LED string is turned on when the corresponding gate signal $\varphi_2$, $\varphi_4$ or $\varphi_6$ is high. The on-time intervals for the three LED strings are denoted as $D_1T_2$, $D_2T_2$ and $D_3T_2$ respectively, which also represent the duration of the on-time intervals. The upper limit for $D_1$, $D_2$ and $D_3$ is 1 so the on-time intervals can be overlapped, as illustrated in Fig.3. One dimming period is equally divided into three sub-intervals of length $T_1 = T_2/3$ and each is allocated to one channel for charging the output capacitor. The charging phases for the three channels are denoted as $\varphi_1$, $\varphi_3$ and $\varphi_5$ with duration $d_1T_2$, $d_2T_2$ and $d_3T_2$, respectively, where $d_1$, $d_2$ and $d_3$ are the duty ratios of the charging phases. The upper limit for $d_1$, $d_2$ and $d_3$ is 1/3. Typically $d_k = D_k/3$ but can be a different value when variable dimming frequency is utilized, as will be explained in Section IV.

For channel 1, the on-time interval $D_1T_2$ has two parts. The first part is $d_1T_2$, during which both $\varphi_1$ and $\varphi_2$ are high, when the boost converter charges the output capacitor and supplies the LED string. In the second part, $\varphi_1$ is low and $\varphi_2$ is high, when the capacitor is disconnected from the boost converter and is the only power supply for the LED string. During the on-time interval $D_1T_2$, $Q_4$ is turned on simultaneously with $Q_3$, thus the LED string and the integrator are turned on and off at the same time. Hence the integrator is called a synchronous integrator. Although the channel voltage is regulated only during the first part of the on-time interval, the powerful and robust synchronous integral control is able to bring it to a steady state value which produces the desired LED current under a wide range of operating conditions. The other two channels have similar working principle.

The key idea of the proposed timing scheme is the separation of the capacitor's charge interval and the LED string's on-time interval for a channel, which is implemented by isolating the switch $Q_2$ from $Q_3$, and that the non-overlapped charging phases along with diode $D_1$ isolate the output capacitors to avoid short circuit among them during $d_1T_2$, $d_2T_2$ and $d_3T_2$. Meanwhile, the on-time intervals $D_1T_2$, $D_2T_2$ and $D_3T_2$ can be overlapped, allowing all LED strings to be turned on at the same time, so that the maximum value for the dimming ratios can be 1 and the LEDs' capacity can be fully utilized. Theoretically, the duty ratio $d_k$ of a charging interval can be one third of the dimming ratio $D_k$. While in practice, to prevent short circuit among output capacitors, dead time logic needs to be implemented between the on times of $\varphi_1$, $\varphi_3$ and $\varphi_5$, which slightly reduces the upper limit of $d_1, d_2$ and $d_3$. Also note that snubber circuits should be used to minimize voltage stress. When no channel capacitor is charged, the power supply, the inductor L, the diode D and the MOSFET $Q_1$ and $Q_2$ are virtually turned off. Therefore, the efficiency should be nearly the same as the case where there is no dimming. By the proposed timing plan, only one capacitor can be charged at a time, but the LED strings can be turned on simultaneously. By the synchronous integral control method, the forward current of the first LED string is regulated during $d_1T_2$ interval, and the output capacitor is discharged during the second part of $D_1T_2$ interval. (Similarly for the other channels.) In case of low dimming frequency, i.e. 300Hz, and high dimming ratio, the output capacitor may be discharged considerably by the end of the $D_1T_2$ interval, and the LED current would deviate significantly from the rated value. To ensure satisfactory steady state performance and acceptable current ripple size, the driver parameters such as $L$, $C$ and dimming frequency need to be carefully chosen based on transient and steady state analysis of the circuit. Transient analysis to be carried out in section III shows that $1.8kHz$ dimming frequency for three LED string driver achieves acceptable results for dimming ratio greater than 30%.

The proposed time multiplexing and control strategies are designed such that independent current regulation of each LED string can be attained within dedicated time interval $d_1T_2, d_2T_2$ or $d_3T_2$. This feature has two advantages. First, bus voltage adjustment is done based on each LED strings characteristics, therefore optimal voltage is applied across each string to ensure minimal power loss. Power consumption in a LED string ($P_{STG}$) can be expressed in terms of LEDs power consumption ($P_{LED}$), switch's power loss ($P_{SWG}$) and current sensor power loss ($P_{CS}$) as below:

$$P_{STG} = P_{LED} + P_{CS} + P_{SWG} \qquad (2)$$

When the dimming switching frequency and LED current are set to desired values, the power losses $P_{CS}$ and $P_{SWG}$ would be constants. Therefore, the string's power consumption varies according to $P_{LED}$.

$$P_{LED} = V_{LED} \times I_{LED} \qquad (3)$$

where

$$V_{LED} = V_{BUS} - \Delta V_{CS} - \Delta V_{SWG} \qquad (4)$$

Since constant LED current and switching frequency lead to constant $\Delta V_{CS}$ and $\Delta V_{SWG}$, the string's power loss is just a function of the bus voltage. This is very important when the driver works in flexible dimming mode and each string has its own dimming ratio. Due to different junction temperature and aging factor, the *I-V* characteristics of the strings would be different and independent current control for each string is necessary to mitigate extra power loss. As investigated in [4], fixed bus voltage for all LED strings can degrade the efficiency up to 30%.

Second, the proposed LED driver offers independent current control of each string, which facilitates flexible dimming scheme to control each individual string by its own dimming ratio. In applications where adaptive dimming technique drives divided units of channel [25, 26], the proposed time multiplexing algorithm can be adopted to increase the accuracy of the image contrast ratio. In [4], a single time-shared regulation loop is implemented by utilizing a single current sensor and minimizing the current balance error among strings. As will be seen in transient analysis to be conducted in Section III, using separate current sensors in the proposed LED driver will not affect the driver's response, since the feedback loop is robust to the change of integrator gain in a wide range.

## III. Transient Analysis and Design Consideration

In this section, a continuous-time mathematical model of the boost converter in PCCM operation is used to analyze the effect of the converter components on the steady state and transient response. This mathematical model is fast and accurate enough to tackle many difficult design trade-offs and also ensures this accuracy at the limits [27]. The mathematical model to be used in this paper is a third order model based on the basic differential equations of the boost converter and control loop, taking all the parasitic series resistors into account. The time-domain solutions of the equations are used to calculate the output voltage, the load current and other circuit variables under certain operating conditions. The equivalent circuit of a boost converter with freewheeling switch is illustrated in Fig. 4. In this equivalent circuit, the parasitic series resistance of inductor $R_L$ and switches' on resistance $R_{Qn}$ are taken into account.

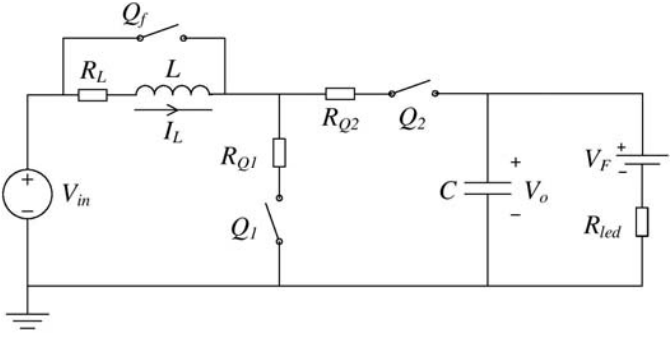

Fig. 4. Circuit diagram of boost converter with parasitic series resistors

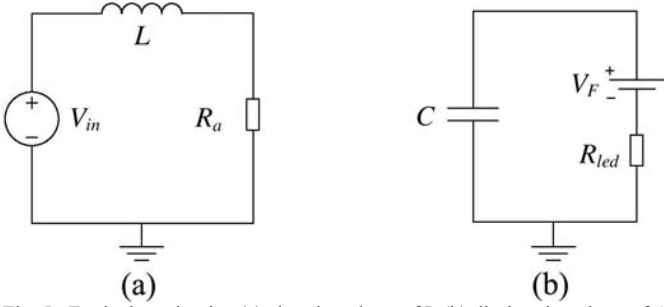

Fig. 5. Equivalent circuits: (a) charging phase of L (b) discharging phase of C

For simplicity, the total equivalent series resistance of the LED string is denoted as $R_{led}$, which includes $R_S$ and the dimming switch's on resistance. The forward voltage drop of the LED string is denoted as $V_F$. The equivalent circuits of the charging phase of $L$, and the discharging phase of $C$ are shown in Fig. 5. Two first order differential equations, together with the initial conditions for $I_L(t)$ and $V_C(t)$, are given as below:

$$L\frac{dI_L}{dt} + R_a I_L = V_{in} \tag{5}$$

$$(CR_{led})\frac{dV_C}{dt} + V_C = V_F \tag{6}$$

with

$$i_L(0) = I_{L\_min} \tag{7}$$

$$V_C(0) = V_{C\_max} \tag{8}$$

$$R_a = R_L + R_{Q1} \tag{9}$$

Where $V_{C\_max}$ is the value of $V_C(t)$ at the end of the preceding charge phase of $C$, and $I_{L\_min}$ is the value of $i_L(t)$ at the end of the preceding discharge phase of $L$ [27]. In PCCM, the converter actually works in DCM in disguise, because the zero dc current in a DCM case is now replaced by a constant $I_{dc}$ [24]. Thus we have:

DCM: $I_{L\_min} = 0 \tag{10}$

CCM: $0 < I_{L\_min} < \infty \tag{11}$

PCCM: $I_{L\_min} = I_{dc} \tag{12}$

The equivalent circuit of the phase when the inductor discharges and the capacitor is charged is depicted in Fig. 6. The second order differential equations for this equivalent circuit, together with two initial conditions, are given as follows:

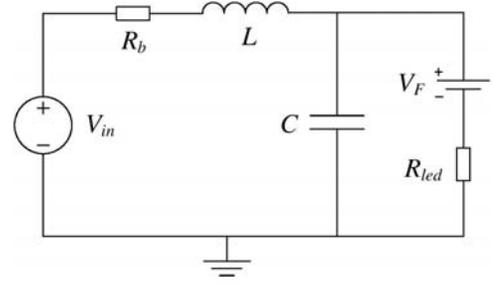

Fig. 6. The equivalent circuit for discharging of L and charging of C

$$\frac{d^2 V_C}{dt^2} + \left(\frac{CR_b R_{led} + L}{LCR_{led}}\right)\frac{dV_C}{dt} + \left(\frac{R_{led} + R_b}{LCR_{led}}\right)V_C$$
$$= \left(\frac{R_{led}V_{in} + R_b V_F}{LCR_{led}}\right) \tag{13}$$

with

$$R_b = R_L + R_{Q2} \tag{14}$$

$$V_C(0) = V_{C\_min} \tag{15}$$

$$\frac{dV_C(0)}{dt} = \frac{1}{C}\left(I_{L\_max} + \frac{V_F - V_{C\_min}}{R_{led}}\right) \tag{16}$$

and

$$\frac{d^2 I_L}{dt^2} + \left(\frac{CR_{led}R_b + L}{LCR_{led}}\right)\frac{dI_L}{dt} + \left(\frac{R_b + R_{led}}{LCR_{led}}\right)I_L$$
$$= \left(\frac{V_{in} - V_F}{LCR_{led}}\right) \tag{17}$$

with

$$I_L(0) = I_{L\_max} \tag{18}$$

$$\frac{dI_L(0)}{dt} = \frac{1}{L}\left((V_{in} - V_{C\_min}) - R_b I_{L\_max}\right) \tag{19}$$

Here $V_{C\_min}$ is the value of $V_C(t)$ at the end of the preceding discharge phase of $C$, which can be expressed as

$$V_{C\_min} = V_F + (V_{C\_max} - V_F)e^{-\frac{t_{on}}{CR_{led}}} \tag{20}$$

Meanwhile, $I_{L\_max}$ is the value of $i_L(t)$ at the end of the preceding charge phase of $L$, which is given by

$$I_{L\_max} = \frac{V_{in}}{R_a} + \left(I_{dc} - \frac{V_{in}}{R_a}\right)e^{-t_{on}\frac{R_a}{L}} \tag{21}$$

Where $t_{on}$ depends on the duty cycle of the boost converter, which is determined by the feedback control law. For the closed loop system with integral control, the duty cycle $D_C$ is approximately a linear function of the integrator voltage ($V_f$). Based on the proposed control strategy, $D_C$ for the first string can be expressed in terms of the closed loop gain $K$ and switching time period of the boost converter ($T_C$), when $Q_4$ is turned on.

$$D_{C\_n+1} = -\left(K/R_{led}\right)\int_0^{T_C}(V_C - V_{Css})\,dt + D_{C_n} \tag{22}$$

where

$$V_{Css} = i_{ref} \times R_{led} + V_F \tag{23}$$

$$T_C = 1/400000 \qquad (24)$$

Since the LED current is a function of the output voltage, to save space, only the output voltage is plotted to examine the effects of design parameters such as $C$, $L$, the integrator gain $K$ and the dimming frequency on the transient responses. The figures shown below are computational results based on the mathematical model for a triple LED strings in parallel and 5 LEDs in each string with rated current of 350mA. The dimming ratio for the string is 95%, which is about the worst case, when maximum capacitor discharging time occurs. Fig.7 shows the effect of $L$ on the transient response. It can be seen that larger inductance reduces overshoot and undershoot, but increases settling time. In case of low dimming ratio, the boost converter has smaller charging time $d_1 T_2$. Thus a driver with larger inductance cannot boost up the output capacitor voltage.

Fig. 8 shows the transient response for several values of the output capacitor. Increasing the capacitance will decrease the voltage overshoot and steady state LED current deviation but slow down the response and increase voltage undershoot.

The transient responses of output voltage for 3 values of $K$ are illustrated in Fig. 9. By increasing the integrator gain $K$, the output voltage oscillation is increased but the settling time is decreased. For K greater than 5000, the closed loop system is unstable. Good transient response is attained for K within [400, 1000]. Larger $K$ results in faster start-up and transient response, thus higher values are preferred, especially when there are more parallel LED strings.

Fig.10 depicts the response of output voltage to the change of the reference current, where the value of the reference LED current is 350mA for $t \in [0,0.4]$; 35mA for $t \in (0.4, 0.8]$ and 350mA for $t > 0.8s$. The voltage response in the figure demonstrates robust stability and desired transient response in following the reference signal. The stability of the closed-loop system is further verified with the Bode plots in Figure 11, for the open loop transfer function obtained at the nominal operating condition (350mA LED current, 7.8V power supply voltage) , which shows an infinity gain margin and a phase margin of 84.1 degree. Within the operating range of 35mA to 350mA LED current and 5.2V to 8.8V power supply voltage, the least phase margin of 68.6 degree is obtained under the operating condition of 35mA LED current and 8.8V power supply voltage, under which the duty ratio for the booster converter main switch is minimal (2.8%).

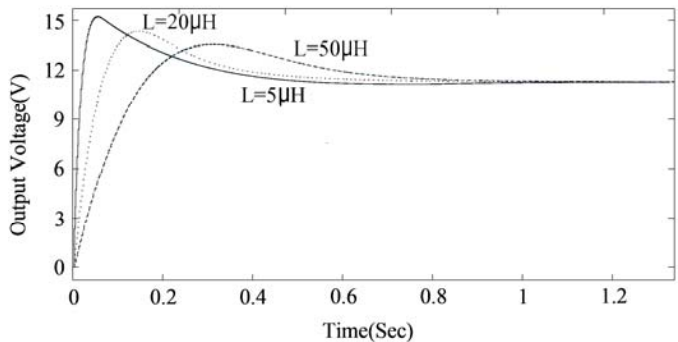

Fig. 7. Transient response of output voltage for different inductance

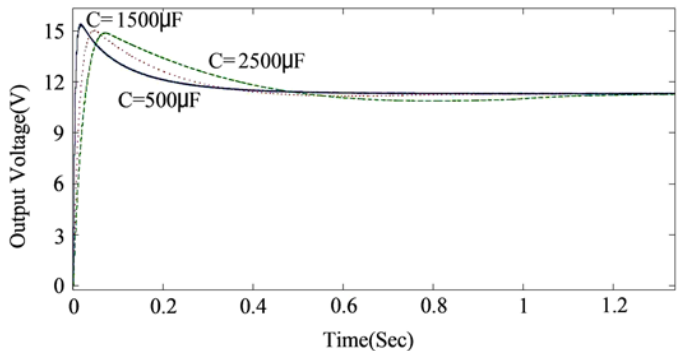

Fig. 8. Output voltage transient response for different capacitance

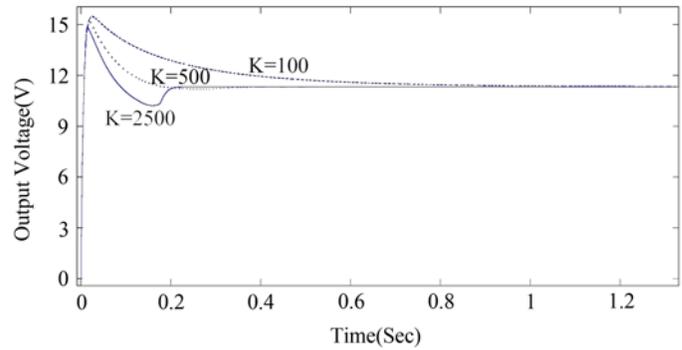

Fig. 9. Output voltage transient response for different values of K

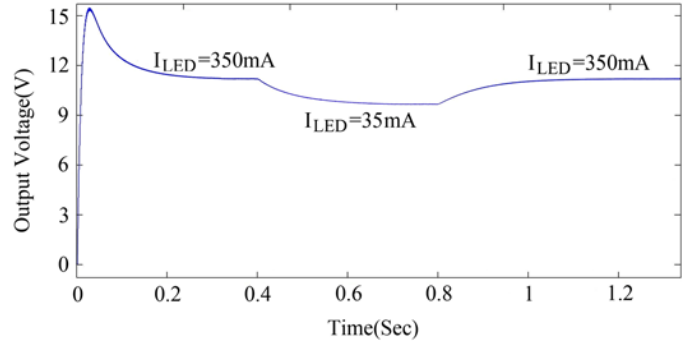

Fig. 10. Output voltage response to the change of reference current

The dimming frequency has a large effect on the steady state deviation of LED current. Lower dimming frequency leads to longer discharging time. Hence, the output capacitor will be more discharged and the LED current will decrease more significantly. On the other hand, higher dimming frequency decreases circuit efficiency, and at low dimming ratio, the output capacitor voltage cannot reach the required steady state. Therefore, based on the mathematical transient analysis, a compromise among response time, overt shoot, undershoot and current deviation must be taken. With regard to the circuit configuration and the parameters of the LED strings, $1/T_2$ =1.8kHz is picked as the dimming frequency, which yields satisfactory transient and steady state response for dimming ratio greater than 30%.

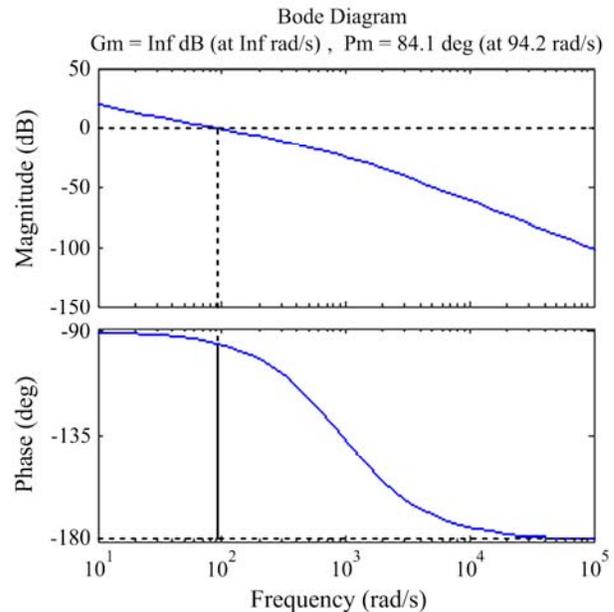

Fig. 11. Bode Diagram under nominal operating condition

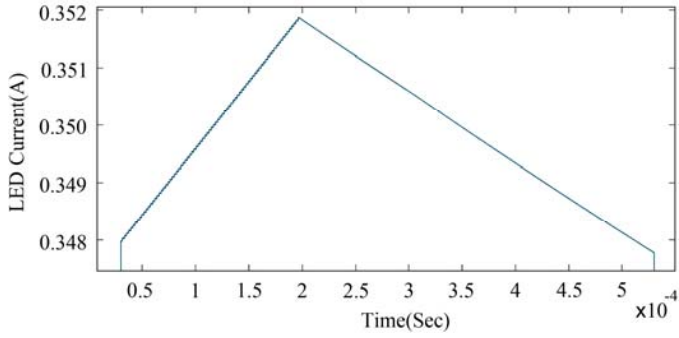

Fig. 12. Transient response of LED current for 1.8kHz dimming frequency and 90% dimming ratio

Fig. 12 shows the transient response of LED current in one period at 1.8kHz dimming frequency. Exponential decaying current response is expected during the discharging phase. With a high dimming frequency, the discharge interval is short and the decrease of the current appears to be a straight line.

The last consideration in the proposed method is the number of parallel strings. Similar issues with regard to scalability has been discussed in [21] for a SIMO buck LED driver. In case of $N$ parallel LED strings, the driver should reach the reference current in $1/N$ dimming period $T_2$ during the capacitor's charging interval and keep the LED current deviation in an acceptable range during the discharging interval of the capacitor, which is $(N-1)/N$ of $T_2$. For a constant boost converter frequency, transient analysis need to be conducted to investigate the trade-off among the circuit parameters such as dimming frequency, integrator gain, inductor and capacitor values, and $I_{dc}$ of PCCM operation. Analysis reveals that decreasing the frequency of the boost converter and increasing $I_{dc}$ can improve the transient response at low dimming ratio, since low frequency allows the capacitor longer time to charge and the boost converter works like a single output converter during the charge interval. Another alternative approach is to adopt supercapacitors which are compatible with the nature of the proposed switching scheme. Supercapacitors can act as output capacitor of boost converter with high frequency switching. On the other hand, it can work as battery in discharging intervals of LED configuration with higher number of parallel strings or series LEDs to maintain the LED current within desired range.

## IV. VARIABLE DIMMING FREQUENCY FOR ENHANCED PERFORMANCE

The mathematical simulation for low dimming ratios shows that, the response time of the boost converter is not fast enough to reach the steady state at high dimming frequency. This imperfection leads to undesirable waveform with considerable undershoot for dimming ratio less than 30%, and makes the LED driver unstable for dimming ratios 15% and lower. Fig. 13 depicts transient responses of output voltage at different dimming ratios when 1.8kHz dimming frequency is implemented. While 90% and 60% dimming ratios yield acceptable responses, lower dimming ratios give slower rise up voltage. The settling time at 25% dimming ratio is 2.62sec, which causes significant LED current undershoots. At 12% dimming ratio, the voltage response gets worse and cannot reach the required steady state value.

To deal with these issues, a variable dimming frequency strategy is devised in the proposed LED driver to achieve satisfactory transient response and nearly square current waveform at steady state for all dimming ratios. With this strategy, the dimming frequency for one string is decreased in 2 steps as a function of the dimming ratio. The function is defined as below:

$$F_D = \begin{cases} F, & D \geq 0.3 \\ F/2, & 0.15 \leq D < 0.3 \\ F/3, & D < 0.15 \end{cases} \qquad (25)$$

Where $F_D$ is the dimming frequency for a LED string, D is the dimming ratio and F is the main dimming frequency, which is selected as 1.8kHz based on the transient analysis for this particular application. To avoid short circuit among LED strings, the main-dimming frequency (1.8kHz) must be divisible by an individual string's dimming frequency. The highest frequency that fulfills the aforementioned condition and yields desired LED current waveform can be picked as the next sub-dimming frequency. This strategy helps to prevent considerable undershoot and reduces the minimum achievable contrast ratio. On the other hand, higher frequency is still adopted for higher dimming ratio to avoid considerable capacitor discharge and maintain the LED current deviation in an acceptable range.

To demonstrate the modified switching signals under the variable dimming frequency scheme, the timing diagram for the three strings with different dimming ratios is presented in Fig. 14, where the dimming ratios are 90%, 25% and 10% for the first, the second and the third string, respectively. The proposed switching strategy is simpler than the counterpart in [28, 29] but more efficient. The corresponding dimming frequency for each LED string is determined as below:

$$F_{D1} = 2F_{D2} = 3F_{D3} = 1/T_{D1} = 1.8kHz \qquad (26)$$

$$F_{D2} = 1/T_{D2} = 900Hz \qquad (27)$$

$$F_{D3} = 1/T_{D3} = 600Hz \qquad (28)$$

where

$$T_{D1} = 3T \qquad (29)$$

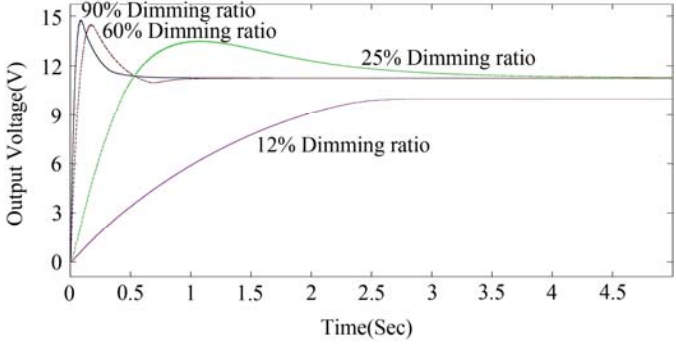

Fig. 13. Output voltage transient response for different dimming ratios

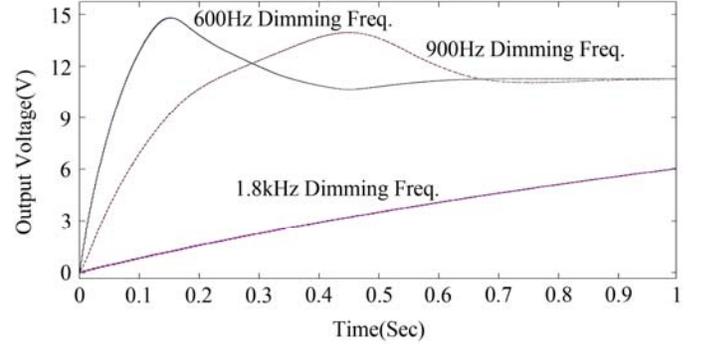

Fig. 15. Output voltage transient responses to 12% dimming ratio

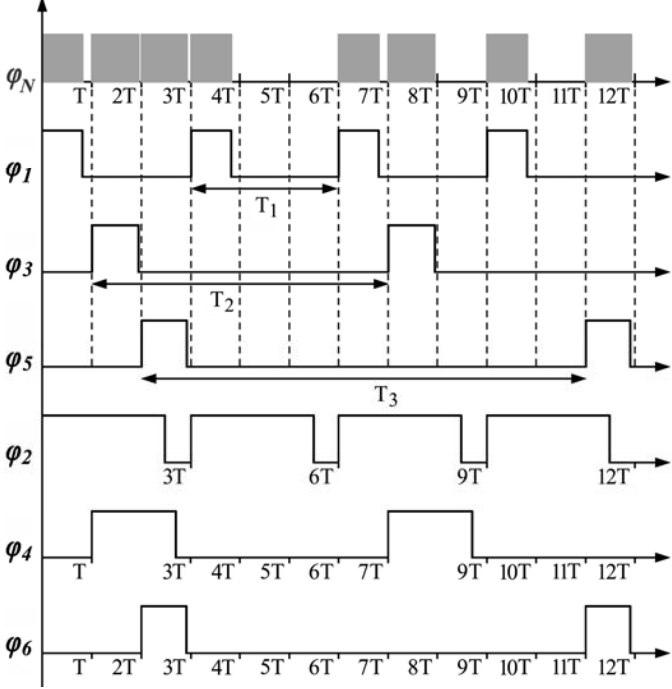

Fig. 14. Modified timing diagram under variable dimming frequency scheme

$$T_{D2}=2T_{D1}=6T \tag{30}$$

$$T_{D3}=3T_{D1}=9T \tag{31}$$

Since the dimming period of the second LED string is selected as $2T_{D1}=6T$, the charging and discharging phases may take T and 5T long, respectively. The dimming frequency corresponding to this dimming period is $900Hz$, which is used when the dimming ratio is between 15% and 30%. This timing scheme has the advantage of using the maximum length of a charging interval, without short circuit among the channels, which is about $T$ for all range of dimming ratios between 15% and 30%. For instance, if the dimming ratio for the second string is 20%, the charging interval would be $T$, the on-time interval would be $0.2 \times 6T = 1.2T$, and the discharge interval when the LED string is solely fed by the capacitor is 0.2T. Note that for low dimming ratios, it is preferred to use maximum charging duty ratio, so that the boost converter has more time to reach the steady state and to yield desired LED current shape. This timing scheme ensures that the charging interval of the second LED string ($\varphi_3$) is ended before the third LED string's charging signal is fired ($\varphi_5$). In case of dimming ratio equal or less than 15%, the LED string works with $3T_{D1}=9T$ dimming period. In this scenario, the maximum charging interval of $T$ is still attainable for all range of dimming ratios equal or less than 15%. For instance, the charging interval for 14% dimming ratio would be $T$, the on-time interval would be $0.14 \times 9T = 1.26T$, and the discharge interval when the LED string is solely fed by the capacitor is 0.26T. On the other hand, for dimming ratios less than 11%, the charging interval could be selected the same as the on-time interval, since the maximum attainable charging interval is greater than the required dimming time. Again, this ensures that the charging interval of the third LED string ($\varphi_5$) is terminated before the first LED string's charging starts ($\varphi_1$), without any short circuit. Fig. 15 shows the effect of reducing dimming frequency on the output voltage's transient response at 12% dimming ratio. A 900Hz dimming frequency improves the transient response, but the settling time of $0.6061Sec$ for 12% dimming ratio is not satisfactory. Further decreasing the dimming frequency to 600Hz yields better transient response and consequently acceptable LED current undershoots. The modified switching scheme allows the LED driver to use different dimming frequencies for each string without switching signal interference, while maintaining the flexible dimming scheme and yields desired waveforms. Higher efficiency is another byproduct of applying the variable dimming frequency scheme. First it reduces the switching loss by applying lower dimming frequency; Second, utilizing maximum attainable charge interval further improves efficiency at lower dimming ratio. It is noted that the efficiency increases as the dimming ratio is increased. Since the average LED current, which contributes to the copper loss, becomes comparable with high peak current of the inductor, which yields constant power loss in all range of dimming ratio when dimming frequency is constant.

## V. EXPERIMENTAL RESULTS

In order to verify the proposed method, a boost converter with a controller based on TMS320F28335 is used to drive three LED strings, where each string consists of 5 LEDs in series. The LED's rated current is 350mA.

Fig. 16 shows the implemented control block diagram by DSP. The outputs $A_0$, $A_1$ and $A_2$ of the analogue to digital converter block in Fig. 16 are the measurements of the three LED strings' currents. The first string's control loop block in

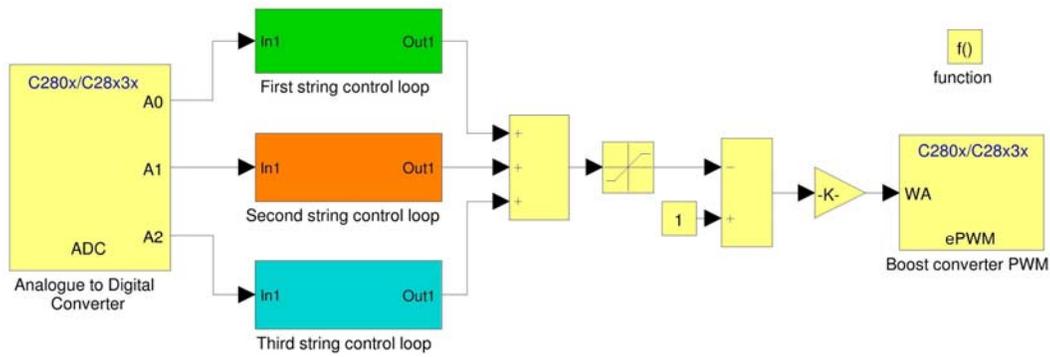

Fig. 16. Experimental control block diagram

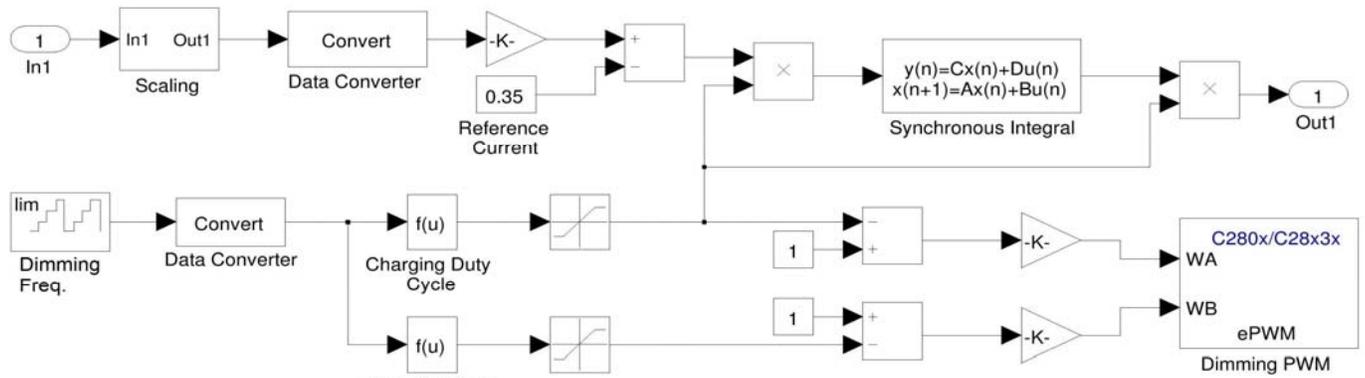

Fig. 17. Expanded block diagram of first string control loop

Fig. 16 is expanded in Fig. 17, where the measured current is compared with the reference value (0.35A) and the error is digitally integrated to generate a duty cycle for the boost converter. The control loop also generates the dimming frequency based on the dimming ratio and switching signals for charging the capacitors and dimming the LEDs. The dimming frequency block in Fig. 17, is a periodic counter based on the frequency of CPU and determines the dimming frequency. The dimming ratio is used to determine the charging duty cycle and dimming duty cycle. The charging duty cycle blocks and the dimming duty cycle blocks for the three strings generate three non-overlapped charging signals, $\varphi_1, \varphi_3, \varphi_5$ and overlapped dimming signals $\varphi_2, \varphi_4, \varphi_6$ respectively. The control loop blocks for other strings are similar to that of the first string, except for that the dimming frequency, charging duty cycle and dimming duty cycle blocks are determined separately based on the dimming ratio of an individual string. The dimming PWM block creates required dimming switching signals for $Q_2$ and $Q_3$. The sum of Out1 from the three control loop blocks is used to generate the intended switching signal for $Q_1$ for the boost converter via the boost converter PWM block in Fig. 16, which is implemented to the MOSFET $Q_1$ via a high frequency MOSFET driver. To ensure safe switching and to prevent short circuit among LED strings, dead time is applied to non-overlapped switching signals and is set to 16μs. Therefore, the duty ratio of charging interval should be kept unchanged for dimming ratios more than 97%.

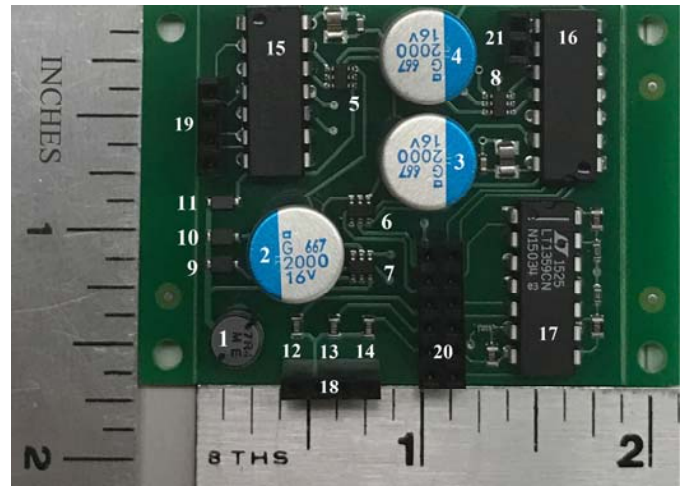

Fig. 18. Photograph of a prototype board

A photo of the prototype board is presented in Fig. 18, where the rulers beside it shows the size of 1.88in by 1.69in. For easy reference, numbers are assigned in Fig. 18 beside the components : #1 -- inductor; #2-4 -- output capacitors; #5-8 -- 8 MOSFETs; #9-11 – diodes; #12-14 -- current sense resistors; #15, 16-- quad MOSFET drivers; #17--quad Op-Amp; #18--power supply, ground and control voltage connectors; #19--connector for switching signals of $Q_1$, $Q_2$, $Q_5$ and $Q_8$; #20 -- connector for LED strings, analogue output signals and switching signals of $Q_3$, $Q_6$ and $Q_9$; #21--connector for switching signal of $Q_f$.

TABLE I
CIRCUIT PARAMETERS

| Symbol | Description | Value |
|--------|-------------|-------|
| $F_C$ | boost converter freq. | 400kHz |
| $F_D$ | main dimming freq. | 1.8kHz |
| $V_{in}$ | dc input voltage | 7.8V |
| $L$ | inductance | 7.4 μH |
| $C_{On}$ | $n^{th}$ output capacitance | 2000 μF |
| $R_{Sn}$ | $n^{th}$ current sense resistor | 0.1 Ω |
| $D_n$ | $n^{th}$ rectifying diode | PMEG10020ELR |
| $Q_n$ | $n^{th}$ MOSFET | FDC6401N |
| - | MOSFET driver | TC4468 |
| - | Op-Amp | LTI1359 |

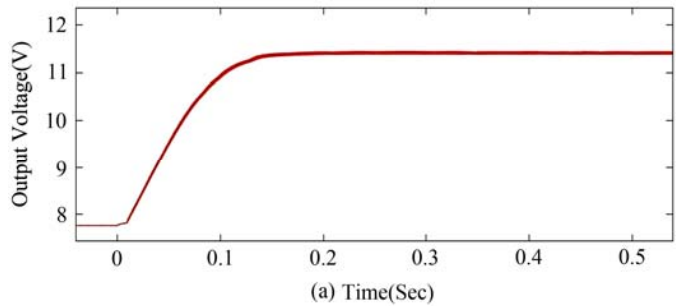

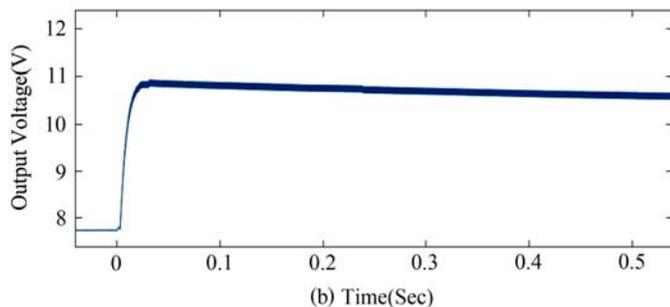

Fig. 19. Output voltage start-up responses (a) 4% dimming ratio (b) 95% dimming ratio

The key parameters of the prototype circuit are provided in Table 1. The steady state waveforms are captured by a 4-channel oscilloscope and the transient responses are recorded by a 24 bit DAQ card (NI 9239), and then plotted by MATLAB.

Figs. 19 and 20 show the start-up responses from experiment, where 4% and 95% dimming ratios are applied to the first and the second string, respectively. Fig. 19(a) shows the start-up response of capacitor voltage for the first string with dimming ratio 4%. The voltage rises up smoothly at a rate about $0.032(V/mS)$ with settling time of 95.83 $mS$, and reaches 11.41V at steady state. The second string's capacitor voltage is plotted in Fig 19.(b), with a 95% dimming ratio. It rises up faster at a rate about $0.209(V/mS)$ and a settling time of 10.93 $mS$. Although the capacitor's voltage reaches about 10.62V at steady state, it continues to decrease due to the current regulation function of the LED driver, in reaction to the increasing junction temperature and the temperature dependent $I$-$V$ curve of the LED string. Similar voltage response has been observed in [30]. For the channel with higher dimming ratio, the output steady state voltage is lower due to higher temperature caused by larger average LED current which reduces the forward voltage or the equivalent series resistance [31]. The difference between the final values of capacitor voltages verifies that individual bus voltage adjustment is attained based on the proposed current regulation control scheme which leads to minimal power loss of the LED strings. The start-up responses of LED currents for channel 1 (4% dimming ratio) and channel 2 (95% dimming ratio) are depicted in Fig. 20. The first LED string with 4% dimming ratio takes $0.168Sec$ to reach the final value of $350mA$, while the second LED string with 95% dimming ratio takes $0.047Sec$. It comes to the conclusion that decreasing the dimming frequency to one third would still result in acceptable transient response of the LED current. Fig. 21 shows the steady state periodic waveforms of LED string currents with 95% uniform dimming ratio for all channels. Note that the LED currents shown in the oscilloscopes are scaled up by 10.

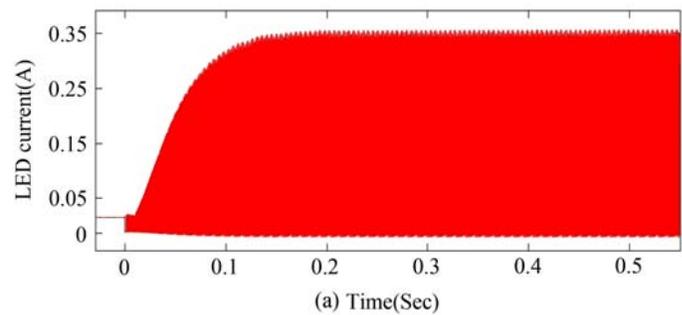

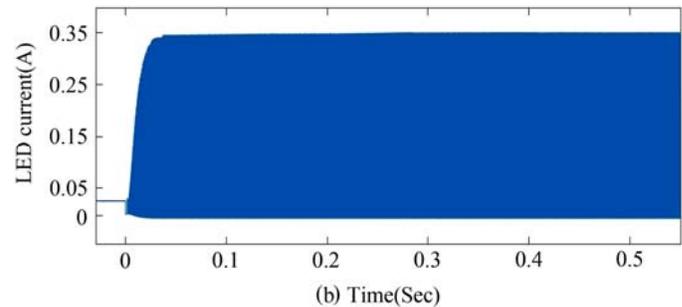

Fig. 20. LED current start-up response for (a) 4% dimming ratio (b) 95% dimming ratio

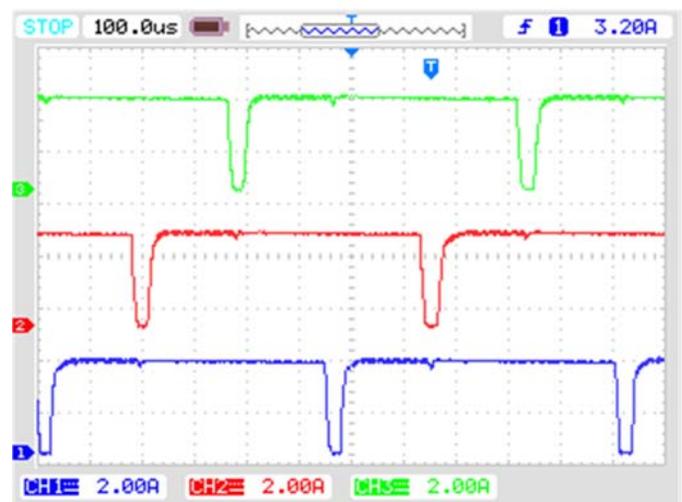

Fig. 21. Steady state results for 95% dimming ratio

TABLE II
CURRENT IMBALANCE AND DEVIATION FOR 95% UNIFORM DIMMING

| Description | Avg. Current(mA) | Current imbalance(%) | Current deviation(%) |
|---|---|---|---|
| *String 1* | 344 | 0.09 | -1.71 |
| *String 2* | 344 | 0.09 | -1.71 |
| *String 3* | 343 | -0.19 | -2.00 |

It can be seen that the on time interval of one string has two parts. The first part is the charging interval, which is marked by larger peak to peak value (thicker line), when the output capacitor is charged by the boost converter. The second part is the discharging interval, which is marked by smaller peak to peak value (thinner line), when the LED string is solely fed by the capacitor. The LED currents slightly increase during charging intervals, which results from the charging current of the boost converter and is an effect of the current regulation of the integral control. While in a discharging interval, since the LED string is solely fed by the capacitor, the capacitor voltage should decrease. However, this decrease is almost invisible in the oscilloscope waveforms owing to the proper choice of switching frequency and boost converter parameters. To measure the current imbalance among the three strings, denote the average of on-time currents as $I_{Avg} = \sum_{n=1}^{3}(1/3)I_n$, where $I_n$ is the average on-time current of the $n$th LED string. Then the current imbalance for each string can be expressed as:

$$\text{Current imbalance} = (I_n - I_{Avg})/I_{Avg} \qquad (32)$$

Table 2 shows the measured average on-time string currents, current imbalance and current deviation from the reference value at 95% uniform dimming ratio. Negligible current imbalance makes the proposed LED driver a perfect match for applications where uniform brightness is required.

Figs. 22 and 23 show the steady state current waveforms at 20% and 4% uniform dimming ratio, where the switching frequency is reduced to 900Hz and 600Hz, respectively, by the variable dimming frequency scheme. These figures validate the desirable performance of the LED driver with the proposed control scheme, and that nearly square current waveforms are generated within a wide range of dimming ratio.

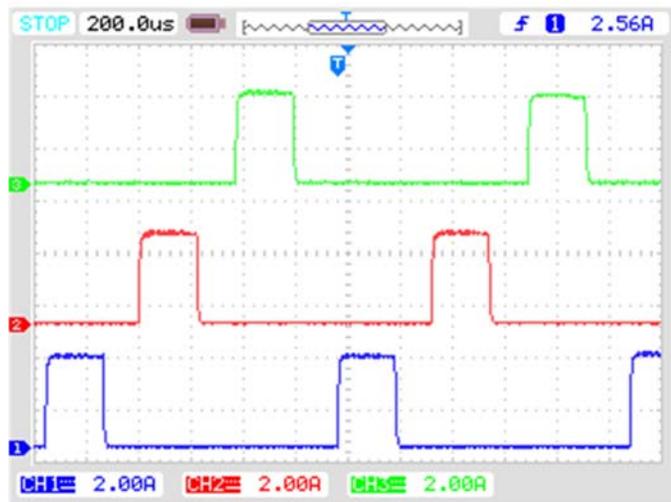

Fig. 22. Steady state results for 20% dimming ratio

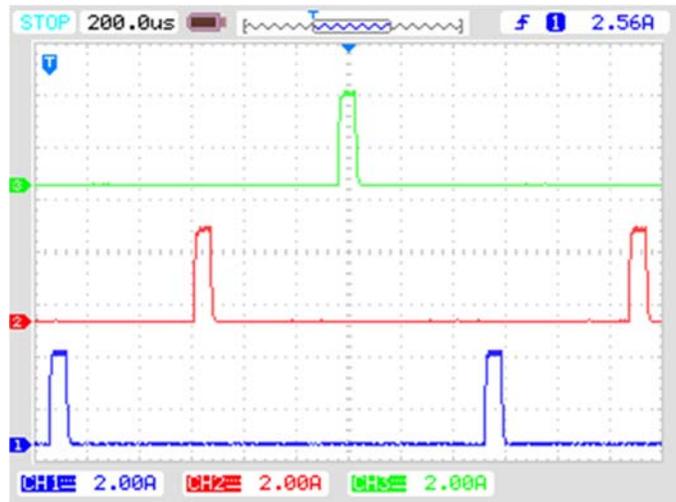

Fig. 23. Steady state results for 4% dimming ratio

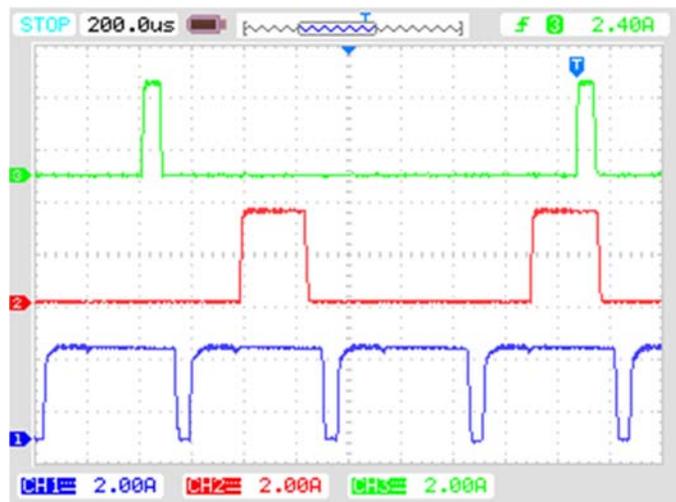

Fig. 24. Steady state results for 90%, 23% and 4% dimming ratio

Fig. 24 demonstrates the flexible dimming capability of the proposed LED driver, where the first string works at 90% dimming ratio (bottom curve), the second string at 23% (middle) and the third string at 4% (top). Table 3 presents the measured average string currents, current imbalance and current deviation from reference value for the three strings with 90%, 23% and 4% dimming ratios.

Fig. 25 is generated to show both the LED current and the bus voltage for the first and the third strings. In this figure, CH1 and CH2 of the oscilloscope (two curves at bottom) represent the LED current and bus voltage of the third string with 4% dimming ratio, and CH3 and CH4 (two curves on top) represent the LED current and bus voltage for the first string at 90% dimming ratio. The bus voltages are adjusted to 11.36V and 10.48V for the third and the first string, respectively. Each bus voltage should increase during the capacitor's charge interval and decrease during its discharge interval. Owing to the proper choice of LED driver parameters and control scheme, there is only slight increase of bus voltage during the charge interval and the decrease during discharge interval is almost invisible. It is also observed that there is visible ripple and noise in the bus voltage during the charge interval, which is caused by high frequency switching of the boost converter. The maximal peak to peak ripple for the bus voltage is less than 0.84V.

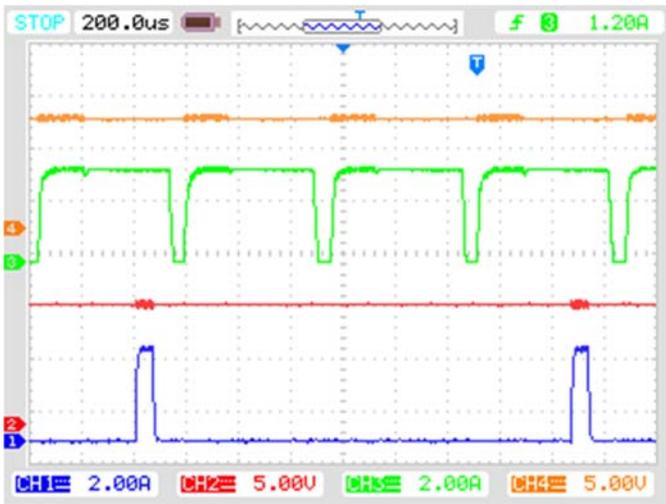

Fig. 25. Steady state results for 90% and 4% dimming ratio

TABLE III
CURRENT IMBALANCE AND DEVIATION FOR VARIABLE DIMMING SCHEME

| Description | Avg. Current(mA) | Current imbalance(%) | Current deviation(%) |
|---|---|---|---|
| *String 1* | 346 | -0.86 | -1.14 |
| *String 2* | 350 | 0.29 | 0 |
| *String 3* | 351 | 0.57 | 0.28 |

Fig. 26 depicts the periodic inductor current waveform (middle curve) at steady state, together with the first LED string's current and bus voltage (bottom and top curves), when the three LED strings work at different dimming ratios, 50%, 95% and 23%, respectively. By the variable dimming frequency scheme, the dimming frequency for the string with 23% dimming ratio is 900Hz while the other two strings' dimming frequencies are 1.8kHz. The inductor current waveform in Fig. 26 shows two charging intervals for the first string (50% dimming), two charging intervals for the second string (95% dimming), and one charging interval for the third string (23% dimming). The inductor current's frequency is 900Hz, the same as that of the string with the lowest dimming ratio.

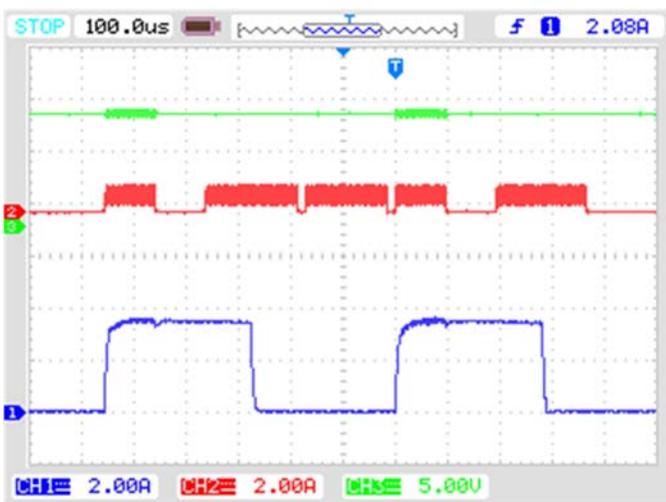

Fig. 26. Steady state results for 50% dimming ratio

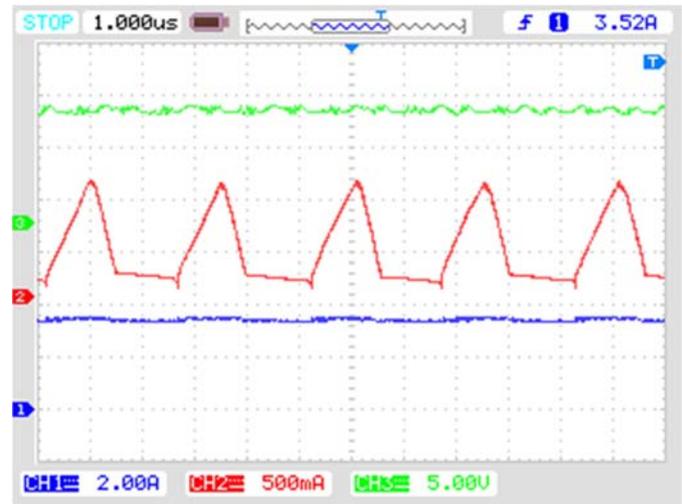

Fig. 27. Close-up view of steady state waveform at 50% dimming ratio

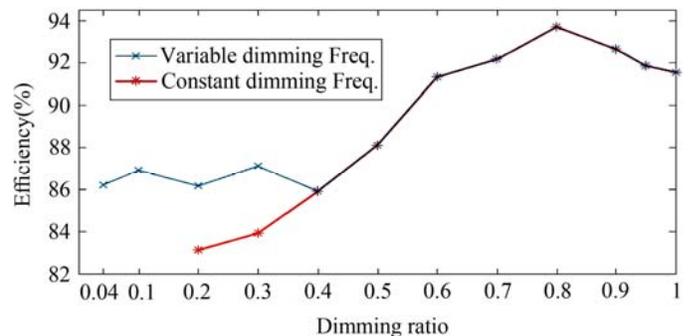

Fig. 28. LED driver efficiency

Fig. 27 is a close-up view of Fig. 26, which shows the waveforms in a few switching periods of the boost converter when one channel's capacitor is charged. The inductor current waveform (the middle curve) indicates that the boost converter operates at PCCM. Note that the inductor current decreases when freewheel switching is activated, since the current loop is not lossless. The parasitic series resistance of L, the on-resistance of $Q_f$, wiring, and bonding are some factors leading to the drop of the inductor current [24]. Fig. 26 shows that flat capacitor voltage during discharging interval of capacitors is attained to recover the LEDs current instantly after the boost converter is turned on for the next dimming cycle. On the other hand, in Fig. 27, PCCM keeps the inductor current nearly constant during off time interval of the boost converter to improve the transient response.

In order to evaluate the efficiency of the proposed LED driver, the *RMS* values of LED currents, voltages across the LED strings, power supply current and voltage are measured. The LED driver's efficiency can be computed as (33),

$$Efficiency = \frac{\sum_{n=1}^{3} V_n I_n}{V_{in} I_{in}} \quad (33)$$

Where $V_n$ is the *RMS* voltage across the $n$th LED string, $I_n$ is the *RMS* current of the $n$th LED string, $V_{in}$ is the power supply voltage and $I_{in}$ is the power supply current. Fig. 28 plots the efficiency versus the dimming ratio, when the same dimming ratio is applied to all LED strings. The maximal efficiency occurred at about 80% dimming ratio. On account of the variable dimming frequency strategy, the efficiency has been improved about 3% at 20% dimming ratio as compared to the constant dimming frequency scheme.

## VI. Conclusion

A novel multiple string LED driver is proposed in this paper by combining synchronous integral control and variable dimming frequency scheme. The driver is highly efficient, has high performance and a compact configuration due to the utilization of a single inductor multiple output boost converter. Independent bus voltage adjustment, which is attained by a novel time multiplexing algorithm and synchronous integrators, minimizes the power loss and improves the efficiency and current balance accuracy. Moreover, the proposed variable dimming frequency scheme allows each string to have independent control of dimming ratio in a wide range, which reduces the minimum achievable contrast ratio to 4%. The *PCCM* operation is adopted in the boost converter to yield fast transient response and very small deviation of output voltages and LED currents. Detailed analysis is conducted on transient performance by using mathematical models. The desirable steady state and transient response performances are verified with a prototype driver circuit for three LED strings, which achieves a current balance error below 1% via the proposed flexible dimming scheme and a dimming range between 4% to 100%.

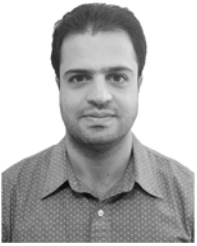

**Mohammad Tahan** (S'15) received the B.Sc. degree in Electrical Engineering in 2006 and his M.Sc. degree in Power electronics from University of Tehran in 2009. Currently he is PhD student at UMass Lowell. His research interests include application of Artificial Intelligence in power electronics, high performance dc/dc converter design, LED driving systems, control applications and battery modeling.

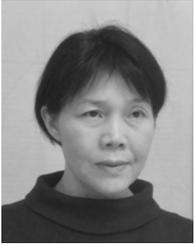

**Tingshu Hu** (SM'01) received the B.S. and M.S. degrees in electrical engineering from Shanghai Jiao Tong University, Shanghai, China, in 1985 and 1988, respectively, and the Ph.D. degree in electrical engineering from the University of Virginia, Charlottesville, in 2001. She was a Postdoctoral Researcher at the University of Virginia and the University of California, Santa Barbara. In January 2005, she joined the Faculty of Electrical and Computer Engineering at the University of Massachusetts Lowell, where she is currently a Professor. Her research interests include nonlinear systems theory, optimization, robust control theory, battery modeling and evaluation, and control applications in power electronics.